\documentclass[conference,10pt]{IEEEtran}
\usepackage{float}

\IEEEoverridecommandlockouts
\usepackage{cite}
\usepackage{amsmath,amssymb,amsfonts}
\usepackage{algorithmic}
\usepackage{graphicx}
\usepackage{textcomp}
\usepackage{xcolor}

\usepackage{tikz}

    \def\Complex{{\rm\rule[.23ex]{.03em}{1.1ex}\kern-.3em{C}}}

    \newcommand{\be}{\begin{equation}} \newcommand{\ee}{\end{equation}}
    \newcommand{\bea}{\begin{eqnarray}} \newcommand{\eea}{\end{eqnarray}}
    \newcommand{\benum}{\begin{enumerate}} \newcommand{\eenum}{\end{enumerate}}

    \newcommand{\qa}{{\bf a}}
    \newcommand{\qb}{{\bf b}}

    \newcommand{\qh}{{\bf h}}

    \newcommand{\qv}{{\bf v}}
    \newcommand{\qw}{{\bf w}}
    
    \newcommand{\qy}{{\bf y}}

     \newcommand{\qphi}{{\boldsymbol \phi}}
    
    \newcommand{\qTheta}{{\boldsymbol \Theta}}

    \newcommand{\qGamma}{{\boldsymbol \Gamma}}
    
    \newcommand{\qOmega}{{\boldsymbol \Omega}}

    \newcommand{\qalpha}{{\boldsymbol \alpha}}

    \newcommand{\qtau}{{\boldsymbol \tau}}

    \newcommand{\qtheta}{{\boldsymbol \theta}}

    \newcommand{\calT}{{\mathcal T}}

    \newcommand{\Ex}{{\tt E}}

    \newcommand*{\argmin}{\operatornamewithlimits{argmin}\limits}

\begin{document}

\title{Labeling Multipath via Reconfigurable Intelligent Surface\\
}
\author{Chang-Jen~Wang$^{\dag}$,~Shang-Ling~Shih$^{\dag}$,~Chao-Kai~Wen$^{\dag}$,~and~Shi Jin$^{\star}$\\
\textit{$^{\dag}$ Institute of Communications Engineering, National Sun Yat-sen University, Taiwan.}\\
\textit{$^{\star}$ National Mobile Communications Research Laboratory, Southeast University, Nanjing, China.}\\
 dkman0988@gmail.com,\ monlylonly@gmail.com,\ chaokai.wen@mail.nsysu.edu.tw, jinshi@seu.edu.cn
}
\maketitle

\begin{abstract}
Reconfigurable intelligent surface (RIS) has shown promise in providing apparent benefits in wireless communication and positioning. Most of the existing research focuses on the ability of RIS to adjust the direction of propagation. In this paper, we present another application based on RIS, named multipath labeling, which intends to inject a label on propagation paths through the RISs.
Each labeled path contains spatial knowledge between the RIS and the receiver, thus opening the
door for sensing the surrounding world by RISs.
The critical challenge is how the labeled paths can be extracted and distinguish from other paths, especially with multipath effects. To address this challenge, we present a complete labeling procedure consisting of channel flipping, parameter extraction, and RIS association. Simulations under a practice ray-tracing model reveal the feasibility of the proposed labeling technique even under the sub 6GHz channels with severe multipath effect. We also apply the multipath labeling technique to localization, which assists user equipment in obtaining its precise location through only a single base station.
\end{abstract}

\section*{I. Introduction}
A vision of future wireless technology is to realize a real-time programmable propagation environment so that we can sense the surrounding world better and improve the quality of wireless communication.
Reconfigurable intelligent surface (RIS) \cite{TWC19-Huang,EURASIP19-Renzo} is gaining popularity as a suitable solution to fulfill such a need owing to properties of easy deployment and low power.
Much research demonstrate that RIS can provide apparent benefits in communication \cite{TSP18-Hu} and positioning \cite{IEEEVTM20-Wym}.

In this study, we present another application based on RIS, named \emph{multipath labeling}.
The multipath labeling technique intends to inject a label on propagation paths, thus opening the door for sensing the surrounding world by RIS.
A high-level principle of this technique operates as follows. A conventional system transmits a modulated signal as usual. Then, an RIS selectively changes the incident signal phase for labeling and scatters the labeled signal. Next, the scattered signal is captured by a receiver and piped through a signal processing engine to extract the label injected by the RIS.
The transmitted signal reflects off multiple objects in the environment and arrives at the receiver as a multipath form.
Only the paths propagating across the RIS are labeled. Therefore, the labeled paths contain spatial knowledge between the transmitter and
the receiver. If the labeled paths can be extracted from multiple paths, we can better sense or control the surrounding world.
For example, if the angle of arrivals (AoA) corresponding to the direct path between the RISs and the receiver can be identified, then the receiver can be localized using simple Triangulation.

Multipath labeling is nothing but slightly perturbing the propagation environment if the labeled paths cannot be extracted, which is the critical challenge.
Specifically, a receiver should receive two types of signals: one directly from the transmitter and one across the RISs. The signal that comes directly from the transmitter is usually much stronger than the scattered signal from the RISs.
Extracting the weak RIS signals from the total revived signals is challenging, especially with the multipath effect.
We leverage an idea called channel flipping \cite{scatterMIMO}. The core idea is shifting the phases of the RISs over time such that the paths across RISs change with time.
As a result, we can separate the RIS signals from the total received signal by removing the direct-current signal over time.
The separated RIS signals remaining suffer from multipath interferences and inter-RIS interferences.
We need to extract the labeling information of RIS signals and associate the signals with corresponding RISs.
This study mainly provides a solution to the multipath labeling technique and evaluates its feasibility under a practice ray-tracing model with a severe multipath effect. We apply the multipath labeling technique to localization, which assist a user equipment to enhance its location precision through only a single base station.

Multipath labeling is closely related to RIS-assisted localization \cite{vtc20-he,iccc20-zhang,ICC20-Denis} and RIS channel estimation \cite{Zheng-20WCL,Noh-20ArXiV}.
However, in these studies, the channel is usually assumed to be a simple two-path channel with a line-of-sight (LoS) path from the transmitter and
a non-line-of-sight (NLoS) path across the RIS. Therefore, most of the methods only can be applied to millimeter wave (mmWave) communications with sparse multipath channels.
To our best knowledge, none of the relevant studies considers the sub 6GHz scenarios, even though the frequency band is the popular 5G and WiFi band. In the sub 6GHz, the two types of signals further involve the multipath effect, which seems to make the position assisted by RIS impossible.
Our proposed multipath labeling can solve the problems in sub 6GHz and can be applied to the mmWave band to simplify the channel model further.
The most relevant work appears to be \cite{scatterMIMO}, in which a particular encoding scheme is designed for RIS to separate the phase of the RIS signal.
However, \cite{scatterMIMO} does not consider path extracting capacity and multiple RISs, and thus contributes nothing relevant to multipath labeling.

\section*{II. Model Setup}
We consider the scenario with a single-antenna transmitter (Tx), $K$ RISs, and a receiver (Rx). Rx is equipped with an antenna array with $N_{\rm R}$ elements.
Each RIS comprises an $M$-element uniform linear array with half-wavelength spacing, and an extension to a uniform planar array is feasible.
We assume that the system is built on OFDM-based transmission, where $s(t,f_{n})$ denote the transmit pilot signal at $f_n$-th subcarrier in $t$-th OFDM symbol. The frequency domain of the received signal can read
\begin{multline}
 \qy(t,f_{n}) = \underbrace{\sum_{l=1}^{L^{[0]}} \alpha_{l}^{[0]} \qa_{\rm R}{\left(\theta_{l}^{[0]} \right)} e^{-j 2 \pi f_n \tau_l^{[0]} } s(t,f_{n}) }_{\rm (1a)\ Tx-to-Rx\ paths}  \\
  +\sum_{k=1}^{K}\underbrace{\sum_{l=1}^{L^{[k]}}
\alpha_{l}^{[k]} \qa_{\rm R}{\left(\theta_{l}^{[k]} \right)} e^{ -j 2 \pi f_n \tau_l^{[k]} }  s(t,f_{n}) }_{\rm (1b)\ Tx-to-RIS-to-Rx\ paths} +\qw(t,f_{n}), \label{eq:channel}
\end{multline}
where parameters $\alpha_{l}^{[\cdot]}$, $\theta_{l}^{[\cdot]}$, $\tau_{l}^{[\cdot]}$ represent the complex coefficient, AoA, and delay of
the $l$-th propagation, respectively. $L^{[\cdot]}$ represents the total number of propagation paths, including LoS and NLoS.
The superscript $^{[k]} $ denotes the parameter associated with the propagation path
across the $k$-th RIS, and $^{[0]} $ denotes those not associated with any RIS. $\qa_{\rm R}(\cdot) \in \mathbb{C}^{N_{\rm R} \times 1}$ is the steering vector for the Rx array, and $\qw(t,f_{n})$ is the additive white Gaussian noise.
We can remove $s(t,f_{n})$ form  \eqref{eq:channel} for ease of expression because the pilot signals are known at Rx.

In \eqref{eq:channel}, the complex coefficient $\alpha_{l}^{[k]}$ for the $k$-th RIS can be further modeled as
\begin{equation}
\alpha_{l}^{[k]} = g_{l}^{[k]} \qa_{\rm RIS}^{T}{\left( \tilde{\theta}_{l}^{[k]}\right)} \qOmega^{[k]} \qa_{\rm RIS}{\left(  \psi_{l}^{[k]} \right)},  \label{eq:alpha}
\end{equation}
where $g_{l}^{[k]}$ denotes the propagation path gain;
$\qa_{\rm RIS}(\cdot) \in \mathbb{C}^{M \times 1}$ is the steering vector for the RIS; $\tilde{\theta}_{l}^{[k]}$ and $\psi_{l}^{[k]}$ denote the incidental and reflection angles of the RIS, respectively.
The matrix $ \qOmega^{[k]}$ is an ${M\times M}$ diagonal matrix of the form
\begin{equation}
\qOmega^{[k]} = {\rm diag}{\left( e^{j \phi_1^{[k]}},\ldots, e^{j \phi_{M}^{[k]}}  \right)},
\end{equation}
where $\phi_{m}^{[k]}$ represents the surface phase for the $m$-th subsurface on the $k$-th RIS. The surface phases of RIS can be controlled and changed in different OFDM symbols. The phase changes make the complex coefficient $\alpha_{l}^{[k]}$ vary with time. To reflect this fact, we replace
\begin{equation}
 \alpha_{l}^{[k]} \rightarrow \alpha_{l}^{[k]}(t)  \label{eq:alpha_t}
\end{equation}
and use $\qphi^{[k]}(t)= ( e^{j \phi_1^{[k]}(t)},\ldots, e^{j \phi_{M}^{[k]}(t)} )$ to denote the phase-shift vector.

A propagation delay $\tau$ manifests the phase shift across the received signal.
Let $\qb(\tau) = \left[ e^{-j 2\pi \tau f_1}, \ldots, e^{-j 2\pi \tau f_{N_s}}\right]^{H} \in \mathbb{C}^{N_s\times 1 } $ be the
phase shift vector due to the delay corresponding to subcarrier $f_1, \ldots, f_{N_s}$.
Then, by staking the received signal and noise vectors with the $N_s$ subcarriers into
\begin{equation}
\qy(t) = {\left[ \begin{array}{c}
\qy(t,f_{1}) \\
 \vdots  \\
 \qy(t,f_{N_s})
\end{array}
\right]} ~{\rm and}~
\qw(t) = {\left[ \begin{array}{c}
\qw(t,f_{1}) \\
 \vdots  \\
 \qw(t,f_{N_s})
\end{array}
\right]},
\end{equation}
we obtain a $(N_{\rm R} N_s)$-dimensional CSI vector
\begin{equation}
\qy(t) = \sum_{k=0}^{K}\sum_{l=1}^{L^{[k]}} \alpha_l^{[k]}(t) \qv{\left( \theta_l^{[k]},\tau_l^{[k]}   \right)} + \qw(t) \label{eq:fre_channel_vec}
\end{equation}
where ${\qv(\theta,\tau) = \qb(\tau) \otimes \qa_{R}(\theta)}$. The $l$-th path associated with the $k$-th RIS can be simply characterized by a 3-tuple $(\alpha_l^{[k]}(t), \theta_l^{[k]}, \tau_l^{[k]} )$, where the surface phases $\qphi^{[k]}(t)$ are included in $\alpha_l^{[k]}(t)$. The model
captures the amplitude and phase changes of the transmit signal, which travels from the Tx to RIS, obtains phase change at RIS, and then travels from RIS to Rx.

\section*{III. Multipath Labeling}

% From an application perspective, once one can obtain the direction of the RIS from at least two RISs (or one Tx and one RIS), the location of Rx can be obtained by using simple triangulation techniques.

The received signal is the superposition of all multipath, including Tx-to-Rx paths and Tx-to-RIS-to-Rx paths.
Our target is to extract the $K$ RIS channels from the received signal and then identify the path that has AoA corresponding
to the LoS path between the RISs and the Rx. Toward this end, we present a complete procedure named multipath labeling, which consists of channel flipping, parameter extraction, and RIS association.

\subsection*{A. Channel Flipping}

First, we express \eqref{eq:fre_channel_vec} as a formulation of \eqref{eq:channel} as
\begin{equation}
\qy(t) =\qh^{[0]} + \sum_{k=1}^{K} \qh^{[k]}(t) + \qw(t)  , \label{eqchannel}
\end{equation}
where
\begin{align}
\qh^{[0]}  &= \sum_{l=1}^{L^{[0]}} \alpha_l^{[0]} \qv{\left( \theta^{[0]},\tau^{[0]}   \right)}, \\
\qh^{[k]}(t)  &= \sum_{l=1}^{L^{[k]}} \alpha_l^{[k]}(t) \qv{\left( \theta^{[k]},\tau^{[k]}   \right)}.
\end{align}
We note from \eqref{eqchannel} that $\qh^{[0]}$ is fixed while $\qh^{[k]}(t)$ is time variant due to the phase toggling by RISs over time slots. Therefore, we can leverage this property to sperate the RIS channels from the overall received signal.

Specifically, by subtracting the received signal vector by its first time slot $t=1$, we obtain
\begin{align}
\Delta \qy_t
    &= \sum_{k=1}^{K} \left(\qh^{[k]}(t)-\qh^{[k]}(1) \right) + \Delta\qw_t, \notag \\
     & =  \sum_{k=1}^{K} \sum_{l=1}^{L^{[k]}} \Delta\alpha_{l,t}^{[k]}  \qv{\left( \theta_l^{[k]},\tau_l^{[k]}   \right)} + \Delta\qw_t, \label{eq:del_y}
\end{align}
for $t=2, \ldots $, where $\Delta\alpha_{l,t}^{[k]} = \alpha_l^{[k]}(t)-\alpha_l^{[k]}(1)$ and $ \Delta\qw_t = \qw(t)-\qw(1)$.
Clearly, only the RIS channels (i.e., $k\neq 0$) remain in $\Delta \qy_t$.
We can set the phase shift to make $\alpha_l^{[k]}(t) = \alpha_l^{[k]}(1) $ for several specific RISs. In this manner, we obtain $\Delta\alpha_{l,t}^{[k]} = 0$ and can further remove inter-RIS interferences from $\Delta \qy_t$.
According to the contains in $\Delta \qy_t$,
we classify the channel flipping mechanism into two modes as follows.
\begin{itemize}
\item {\bf Multilabeling mode}. In this mode, $\Delta \qy_t$ comprises multiple RIS channels as in \eqref{eq:del_y}. This mode provides multiple diversities for RIS channels, but it complicates the subsequent parameter extraction procedure.

\item {\bf Single-labeling mode}. In this mode,
$\Delta \qy_t$ only contains a single RIS channel.
This mode can be realized by changing the surface phases of only one RIS at each time slot while other RISs keep the same phase.
For ease of understanding, we take an example with two RISs and three time slots.
In the first time slot, the two RISs set their initial surface phases. In the second time slot, RIS1 varies the surface phase while RIS2 keeps the same phase as in the first time slot.
In the third time slot,  RIS2 varies the surface phase while RIS1 keeps the same phase as the first time slot.
Therefore, we can resolve RIS1 channel and RIS2 channel from  $\Delta \qy_2$ and $\Delta \qy_3$, respectively.

\end{itemize}
\subsection*{B. Parameter Extraction}
The RIS channels in $\Delta \qy_t$ are usually weak, while we can improve the channel quality by using the multiple snapshots of $\Delta \qy_t$ to mitigate the noise effect.
To extract the path parameters (i.e., delay, AoA, and complex gain) of the RIS channel, we apply multisnapshot Newtonized orthogonal matching pursuit (MNOMP) \cite{SP19-Zhu}. The NOMP-based method \cite{TSP16-Mam} exhibits better performance than other state-of-the-art algorithms with a lower complexity.

For ease of notation, we first consider the single-labeling mode and remove the superscript $[k]$ from \eqref{eq:del_y}. In this case, we rewrite \eqref{eq:del_y} as

\begin{equation}
\Delta \qy_t   = \sum_{l=1}^{L} \Delta\alpha_{l,t}   \qv{\left( \theta_l ,\tau_l   \right)} + \Delta\qw_t, \label{eq:del_y2}
\end{equation}
for $t \in {\calT = \{ 2, \ldots, T \}}$, where $T$ is the number of snapshots.
The problem of extracting 3-tuples of the paths
$( \Delta\qalpha, \qtheta, \qtau ) = (\{\Delta\alpha_{l,2}, \ldots,  \Delta\alpha_{l,T}\}, \theta_l,  \tau_l) $ can be written as the joint minimization problem
\begin{equation}\label{eq:opt_problem}
\argmin_{\Delta\qalpha, \qtheta, \qtau} \sum_{t=2}^{T} \Big\|  \Delta \qy_t - \sum_{l=1}^{L} \Delta\alpha_{l,t}   \qv{\left( \theta_l ,\tau_l   \right)}  \Big\|_2^2.
\end{equation}

Optimization problem \eqref{eq:opt_problem} involves high-dimensional jointly searching of $( \Delta\qalpha, \qtheta, \qtau )$, which is intractable.
MNOMP uses a detection-estimation method to separate each path iteratively and solve \eqref{eq:opt_problem} efficiently.
MNOMP first identifies the strongest signal path, subtracts it from $\left\{ \Delta\qy_{t} \right\}_{t \in \calT} $, and then determines the next strongest path using the residual signal.
The procedure consists of two stages:
\begin{itemize}
\item [1)] Coarse detection. Coarse estimations of AoA, ToA, and complex gain are obtained by precomputed $\qv( \theta, \tau )$ as follows

\begin{subequations}
\begin{equation}
( \hat{\theta}, \hat{\tau} )=\underset{\theta \in \qTheta, \tau \in \qGamma}{\rm argmax}\ \sum_{t=1}^{T}| \qv^{H}( \theta, \tau ) \Delta \qy_t |, \label{eq:maxfunction}
\end{equation}
\begin{equation}
\Delta\hat{\alpha}_t = { \qv^{H} ( \hat{\theta}, \hat{\tau} ) \Delta\qy_t }/{ \| \qv( \hat{\theta}, \hat{\tau} )\|_2^2 },\ t \in \calT,
\end{equation}
\end{subequations}
where $\qTheta$ and $\qGamma$ are the discretized sets of AoAs and ToAs, respectively.

\item [2)] Refinement. Let
\begin{equation}\label{eq:rcost}
    J_{\rm r}(\Delta\qalpha, \theta, \tau) = \sum_{t=2}^{ T} \left\|  \Delta \qy_t - \Delta\alpha_{t}   \qv{\left( \theta ,\tau   \right)}  \right\|_2^2,
\end{equation}
where $\Delta\qalpha = (\Delta\alpha_{2},\ldots,\Delta\alpha_{T})$.
The estimations are refined using the Newton method:
\begin{subequations}
\begin{align}
( \hat{\theta}, \hat{\tau} ) &\leftarrow ( \hat{\theta}, \hat{\tau} ) -  [ \ddot{J_{\rm r}}(\hat{\qalpha}, \hat{\theta}, \hat{\tau}) ]^{-1} \dot{J_{\rm r}}(\hat{\qalpha}, \hat{\theta}, \hat{\tau}),\\
\Delta\hat{\alpha}_t &\leftarrow  {  \qv^{H} ( \hat{\theta}, \hat{\tau} ) \Delta\qy_t }/{ \| \qv( \hat{\theta}, \hat{\tau} )\|_2^2 },\ t \in \calT, \label{eq:est_alpha}
\end{align}
\end{subequations}
where $\ddot{J_{\rm r}}(\cdot)$ and $\dot{J_{\rm r}}(\cdot)$ are the Hessian and gradient of $J_{\rm r}$ with respect to $(\theta, \tau)$ at the current estimation
$(\hat{\theta}, \hat{\tau})$ provided by \eqref{eq:maxfunction}, respectively.
For $t \in \calT$, we subtract the detected path from $\Delta\qy_{t} $ and obtain the corresponding residual signal $\Delta\qy_t \leftarrow \Delta \qy_t - \Delta\hat{\alpha}_t\qv(\hat{\theta},\hat{\tau}) $.
\end{itemize}
The two steps are repeated with the residual signal to estimate other paths. To improve accuracy, the refinement steps are
repeated after each new detection for all the paths in a cyclic manner for a few rounds.

\subsection*{C. RIS Association}
We notice that the complex coefficient $\hat{\alpha}_t$ contains the labeling information of a RIS, which enables us to associate the path with its labeling RIS.
To make the idea explicitly, we unfold the complex coefficient of RIS $\qa_{\rm RIS}^{T}{( \tilde{\theta}_{l}^{[k]})} {\rm diag}( \qphi^{[k]}(t) ) \qa_{\rm RIS}{(  \psi_{l}^{[k]} )}$ as
\begin{equation}
   \sum_{m=1}^{M} e^{  j a_m(\tilde{\theta}_{l}^{[k]}) + j  \phi_m^{[k]}(t) + j  a_m(\psi_{l}^{[k]})     }, \label{eq:RIS_phase}
\end{equation}
where $a_m( \cdot)$ denotes the array response of the $m$-th subsurface of RIS. If we fix the phases of all subsurfaces, said $\phi_m^{[k]}(t) = \phi^{[k]}(t)$ for $m=1, \ldots,M$, then \eqref{eq:RIS_phase} can be written as
\begin{equation}
   \qa_{\rm RIS}^{T}{( \tilde{\theta}_{l}^{[k]})} {\rm diag}( \qphi^{[k]}(t) ) \qa_{\rm RIS}{(  \psi_{l}^{[k]} )} = a_{l}^{[k]} \gamma_{l}^{[k]}(t), \label{eq:RIS_phase2}
\end{equation}
where
\begin{equation}
a_{l}^{[k]} = \sum_{m=1}^{M} e^{j a_m(\tilde{\theta}_{l}^{[k]})+j a_m(\psi_{l}^{[k]}) }, ~~ \gamma_{l}^{[k]}(t) = e^{j \phi^{[k]}(t)}.
\end{equation}
By combining \eqref{eq:RIS_phase2} and \eqref{eq:alpha}, we write \eqref{eq:alpha_t} as
\begin{equation}
\alpha_{l}^{[k]}(t) = g_{l}^{[k]} a_{l}^{[k]} \gamma_{l}^{[k]}(t) = \beta_{l}^{[k]} \gamma_{l}^{[k]}(t), \label{eq:alpha_withGamma}
\end{equation}
where we define $\beta_{l}^{[k]} = g_{l}^{[k]} a_{l}^{[k]}$ for ease of expression.
Applying \eqref{eq:alpha_withGamma}, we write
\begin{equation} \label{eq:D_alpha_lk}
\Delta\alpha_{l,t}^{[k]} = \alpha_{l}^{[k]}(t)  - \alpha_{l}^{[k]}(1)  = \beta_{l}^{[k]} \Delta\gamma_{l,t}^{[k]},
\end{equation}
where $\Delta\gamma_{l,t}^{[k]} = \gamma_{l}^{[k]}(t)-\gamma_{l}^{[k]}(1) $. In \eqref{eq:D_alpha_lk}, $\beta_{l}^{[k]}$ is the coefficient depending on the propagation attenuation and incidental/reflection angles, whereas $\Delta\gamma_{l,t}^{[k]}$ is the coefficient depending on the surface phases of the RIS.

For the single-labeling mode, the $t$-th snapshot for the $k$-th RIS reads
\begin{equation}
\Delta \qy_t  = \sum_{l=1}^{L^{[k]}} \Delta\alpha_{l,t}^{[k]}  \qv{\left( \theta_l^{[k]},\tau_l^{[k]}   \right)} + \Delta\qw_t, \label{eq:del_yt_single}
\end{equation}
where $t$ belongs to certain time slots ${\cal T}_k$. After extracting path parameters by MNOMP, we obtain complex coefficients $\{ \Delta\hat{\alpha}_{l,t}^{[k]}: l =1, \ldots, L^{[k]}, t \in {\cal T}_k \}  $.
We can extract the propagation attenuation coefficient of the $k$-th RIS with multiple snapshots via
\begin{equation}
 \hat{\beta}_{l}^{[k]} =  \frac{1}{|{\cal T}_k|}\sum_{t \in {\cal T}_k } \frac{ \Delta\hat{\alpha}_{l,t}^{[k]} }{ \Delta\gamma_{l,t}^{[k]} }.
\end{equation}
Then, the LoS parameter between the $k$-th RISs and the Rx can be determined by selecting the maximum gain of $ \hat{\beta}_{l}^{[k]}$ from $ l =1, \ldots, L^{[k]} $.

For the multilabeling mode, we also can obtain complex coefficients $\{ \Delta\hat{\alpha}_{l,t} : l = 1, \ldots, \sum_{k=1}^{K} L^{[k]}, t \in \calT \}$ by using MNOMP.
The extracted complex coefficients are not associated with any RISs in this stage. We notice that if a path is belong to the $k$-th RIS, then
\begin{equation}
  \frac{\Delta\alpha_{l,t}}{\Delta\alpha_{l,t+1}}
  = \frac{\Delta\gamma_{l,t}^{[k]}}{ \Delta\gamma_{l,t+1}^{[k]}} = \frac{    e^{j \phi^{[k]}(t)}     -e^{j \phi^{[k]}(1)} }{  e^{j \phi^{[k]}(t+1)} - e^{j \phi^{[k]}(1)}}.
\end{equation}
Therefore, we calculate $\Delta\hat{\alpha}_{l,t}/\Delta\hat{\alpha}_{l,t+1}$ and then use it to associate the RIS.
We use multiple snapshots of phase difference $\{\Delta\hat{\alpha}_{l,t}/\Delta\hat{\alpha}_{l,t+1} \}_{t=2}^{T-1}$
to improve the identification reliable because a snapshot of phase difference is unreliable when noise interference exists.
\begin{figure}[!htp]
\setlength{\abovecaptionskip}{-1.0pt}
\setlength{\belowcaptionskip}{4.0pt}
\begin{center}
\resizebox{3.0in}{!}{%
\includegraphics*{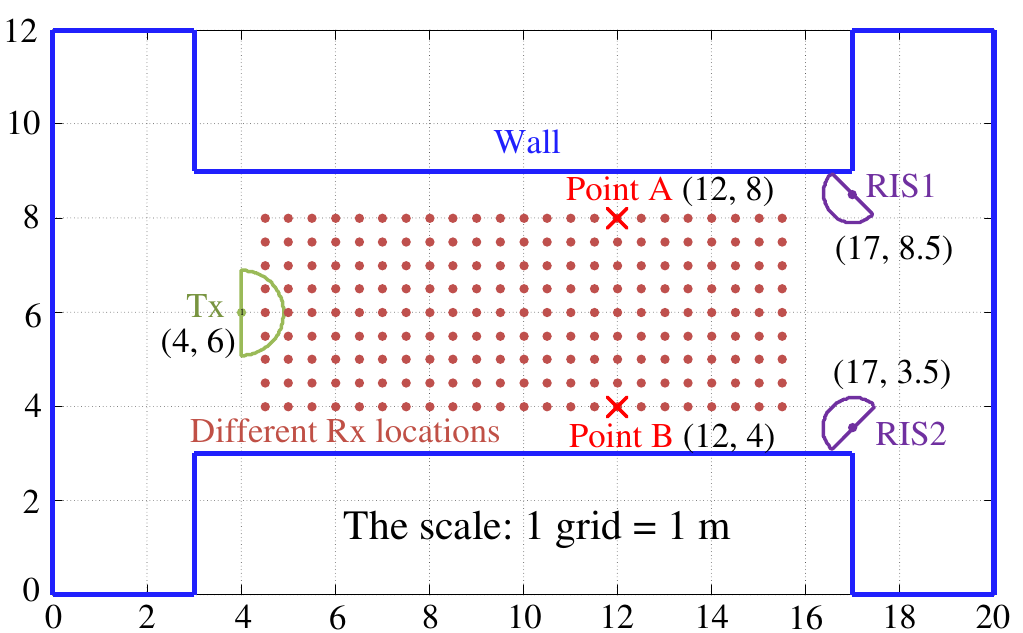} }%
\caption{A floor plan with Tx (denoted by green circle), 2 RISs (denoted by purple circle), different locations of Rx (denoted by red circle), and wall reflections.}\label{fig:Fig-SS}
\end{center}
\end{figure}

\section*{IV. Numerical Results}
We conduct simulations to evaluate the proposed technique in single and multiple labeling modes.
The channel is generated by a ray-tracing simulator.

%\footnote{The simulator consists of many access points and users geographically distributed in a certain outdoor or indoor environment. The outputs of the ray-tracing simulator include the channel parameters (angles or arrival/departure, path gains, etc.) for the channels between every transmitter and receiver.}
\subsection*{A. Simulation Setup}
For the simulator, we consider an OFDM system following the 5G NR frame structure \cite[ch. 5.4]{5GNR}. The system operates at $3.5$ GHz with $100$ MHz bandwidth and $2048$ subcarriers, where each subcarrier spaces $60$ kHz.
The indoor layout is depicted in Fig.\,\ref{fig:Fig-SS}. We place the Tx and $2$ RISs in fixed locations while we vary the Rx locations. The transmission power of Tx is $1$ mW.
Rx is equipped with a triangular antenna array, and the distance between the three antennas is half wavelength.
Unlike the uniform linear array, which has ambiguity in the front and back directions, the triangular array used here can estimate AoA  with $360^{\circ}$. The RIS is equipped with $M=10$ subsurfaces with fixed orientation, as shown in Fig. \ref{fig:Fig-SS}.

For Tx-to-Rx components, we consider one LoS and many NLoS paths, which are made up of one, double, and triple reflections from nearby walls
\cite[ch. 2]{Ray-tracing}.
For Tx-to-RIS-to-Rx components, we consider a direct path from the Tx to the RIS,\footnote{We also can consider reflection paths from the Tx to the RIS.} and then the RIS scatters the incident signal from the Tx. The scattered signals reflect off multiple objects and become multiple paths arriving at the Rx. The paths from the RIS to the Rx also include one LoS and many one-to-triple reflections.
The power of the propagation coefficient is modeled by $| g_{l}^{[k]}| = 10^{ P^{[k]} /20} \times 10^{P_l^{[k]} /20}$ \cite{Ozdogan-20WCL},
where the path losses of Tx-to-RIS $P^{[k]}$ and RIS-to-Rx $P_l^{[k]}$ are derived using the standard free-space propagation model.
The RIS-to-Rx path considers the reflection path, where each has a different propagation distance, such that its path loss is indicated by the subscript ``$_l$'' to reflect this effect.
In addition, we assume the addition of $5$\,dB attenuation per reflection.

For the $t$-th time slot, the phase shift of RIS1 and RIS2 are denoted by $(\phi^{[1]}(t),\phi^{[2]}(t))$.
The received signals adds white Gaussian noise at the level of $10^{-7}$, which results in the measured signal-to-noise ratio of $25$ dB at the central point $(10, 6)$ in Fig.\,\ref{fig:Fig-SS}. To indicate the performance of multipath labeling, we use the mean absolute error (MAE) of AoA
\begin{equation}
    {\rm MAE} = \Ex\left\{ \left|\theta_1^{[k]} - \hat{\theta}_1^{[k]} \right| \right\}, 
\end{equation}
where $\theta_1^{[k]}$ and $\hat{\theta}_1^{[k]}$ are the true and the estimated AoAs of the LoS path between the $k$-th RIS and the Rx.
We obtain each subsequent result through the average of $1,000$ Monte Carlo trials.
\begin{figure}[!htp]
\setlength{\abovecaptionskip}{-1.0pt}
\begin{center}
\resizebox{2.7 in}{!}{%
\includegraphics*{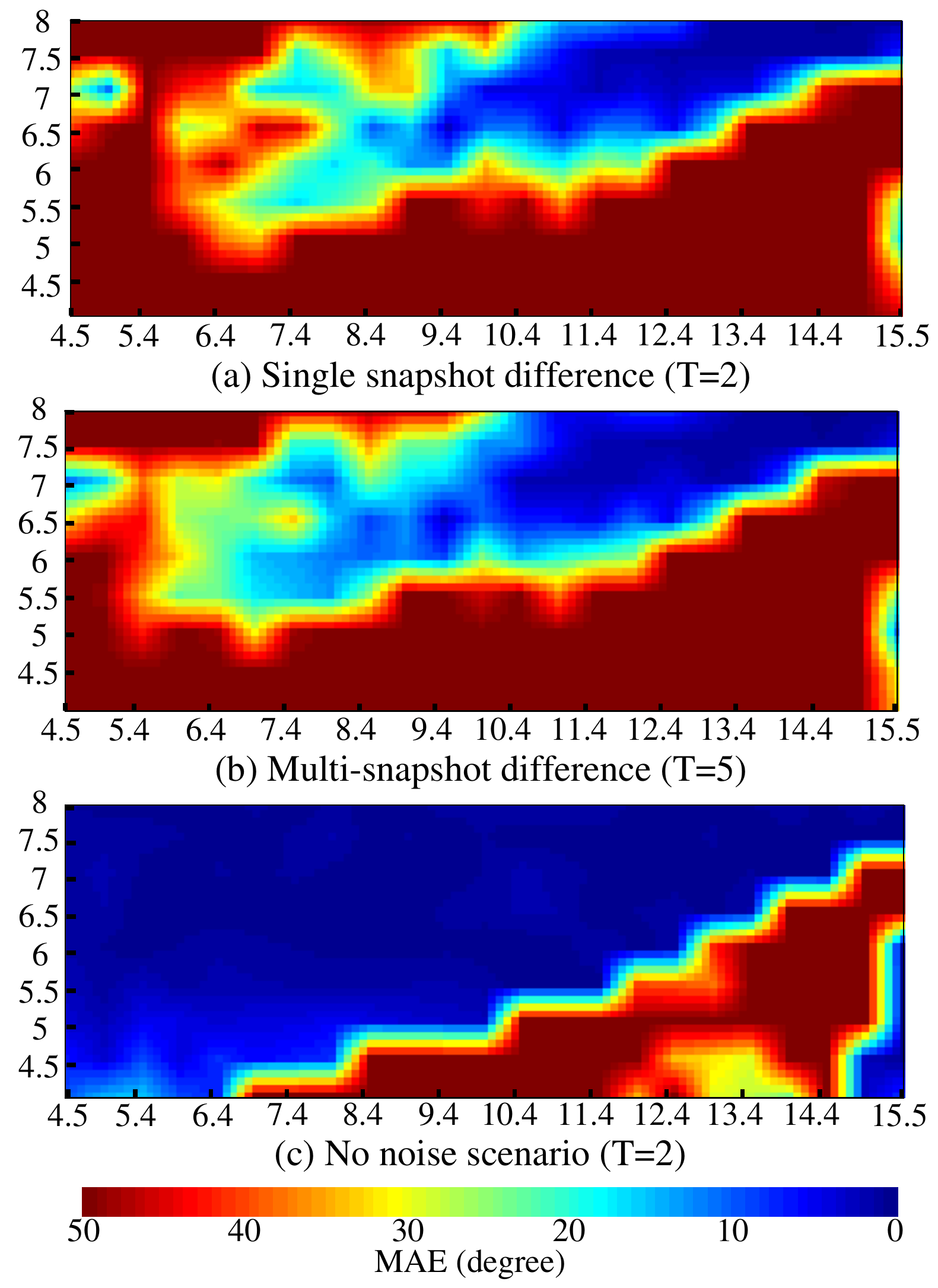} }%
\caption{MAE of AoA under single labeling RIS1 channel for different location of Rx under (a) single time slot difference, (b) multiple time slot difference, (c) no noise interference.}\label{fig:Fig-orgAoA}
\end{center}
\begin{center}
\resizebox{3.5 in}{!}{%
\includegraphics*{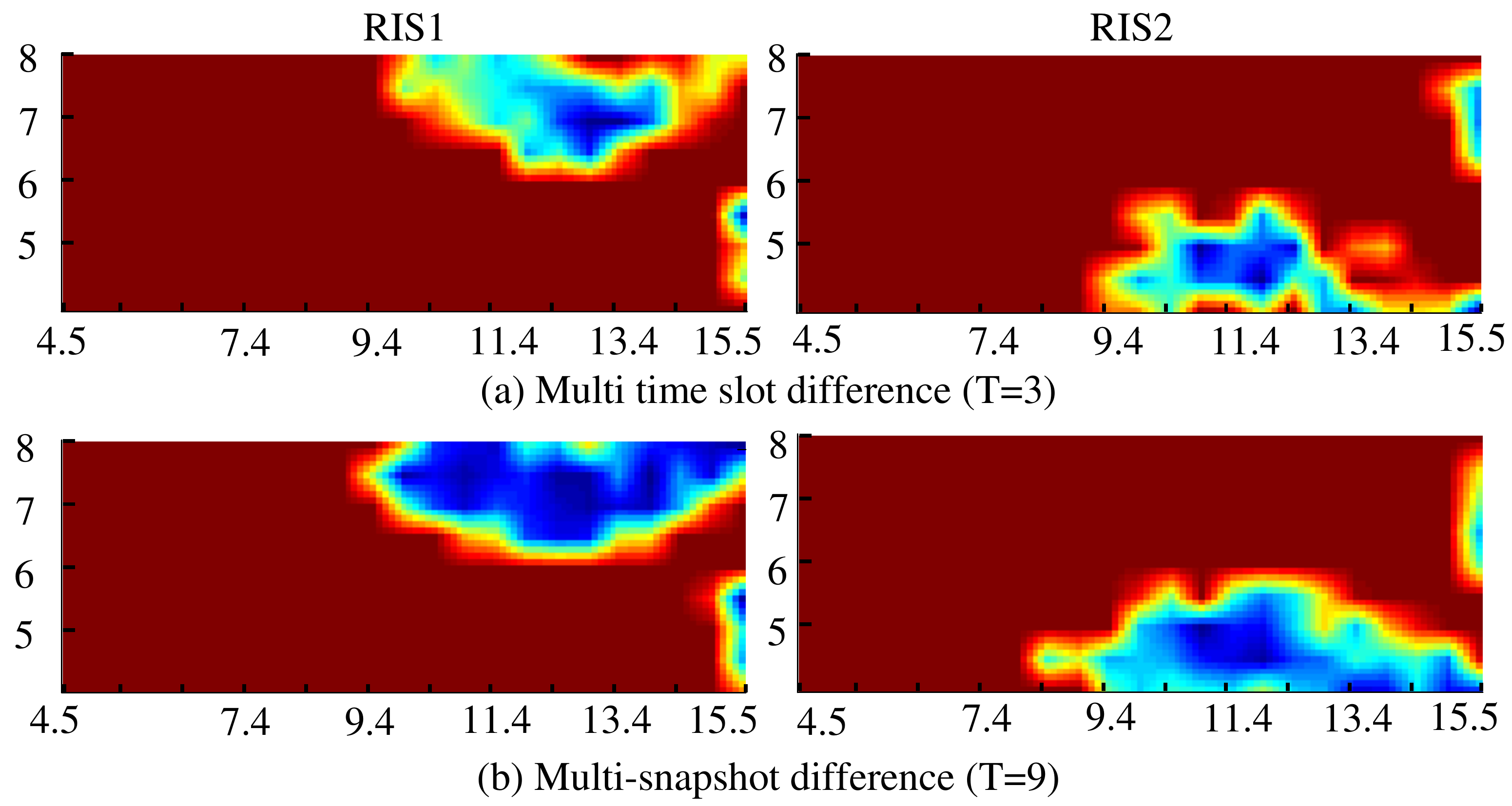} }%
\caption{MAE of AoA under multilabeling RIS1 and RIS2 channels for different locations of Rx under (a) lowest requirement time slot difference and  (b) sufficient time slot difference.}\label{fig:non-orgAoA}
\end{center}
\begin{center}
\resizebox{3.2 in}{!}{%
\includegraphics*{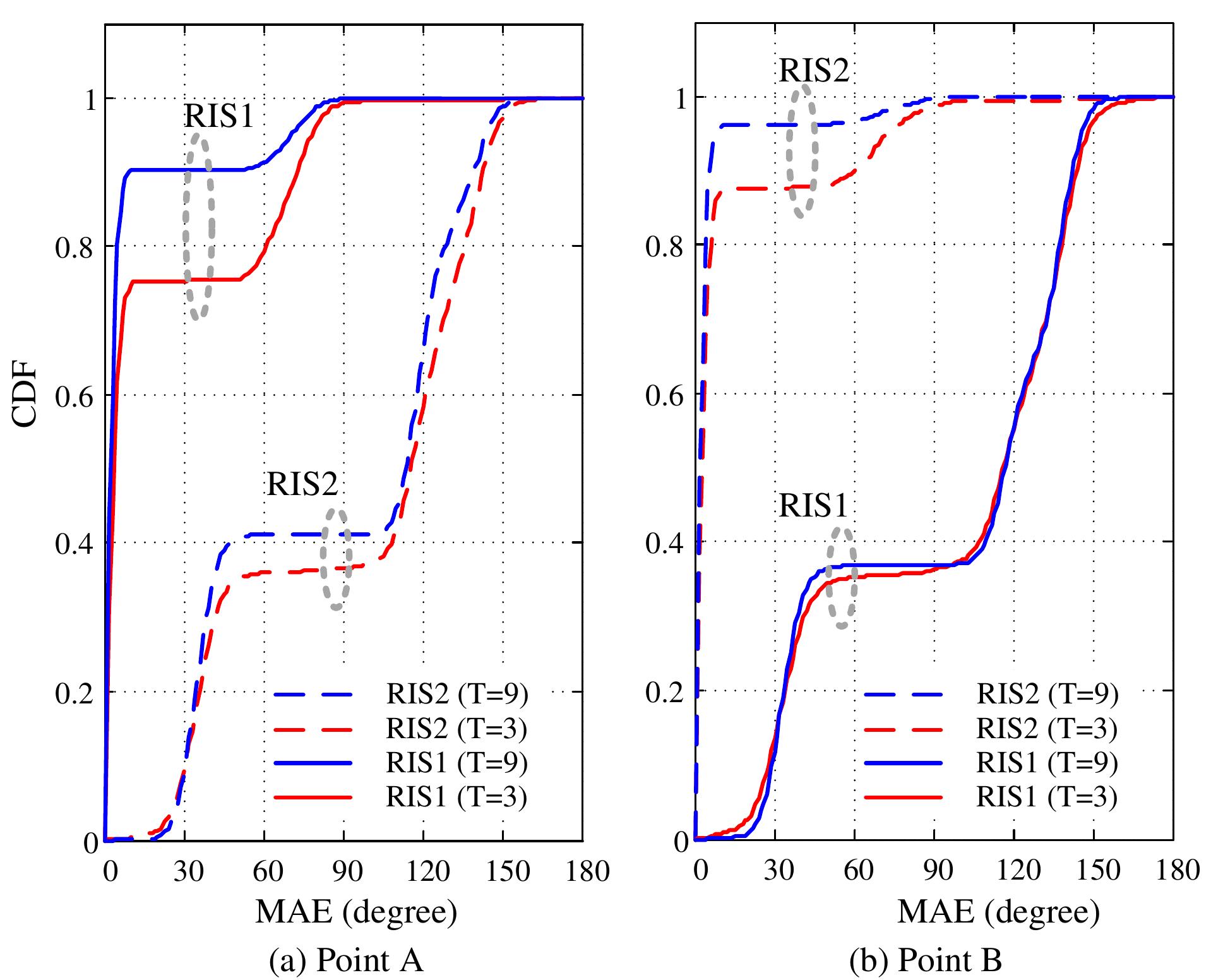} }%
\caption{MAE of RIS1 and RIS2 under two locations.}\label{fig:non-orgAoA-CDF}
\end{center}
\end{figure}

\subsection*{B. AoA Estimation}

Fig.\,\ref{fig:Fig-orgAoA} shows the MAE of the AoA estimation for single-labeling mode under different number of snapshots.
For $T=2$, we set ${(\phi^{[1]}(1),\phi^{[2]}(1))}  = (0,0)$ and ${(\phi^{[1]}(2),\phi^{[2]}(2))} = (\pi,0)$.
With this setting, we can separate the labeled multipath of RIS1 from others in $\Delta \qy_{2}$.
Fig.\,\ref{fig:Fig-orgAoA}(a) illustrates the MAE with the single snapshot observation.
The low MAE area (the blue area) is the location near RIS1. MAE degenerates when Rx is away from RIS1, so the shape of the blue area is related to the scattering energy of RIS1.
Several areas are close to RIS1 but have large MAE because the area suffers from serious multipath interference resulting from the RIS scattering.
In this situation, the LoS of RIS1 is difficult to be detected.

To understand the influence of noise and multipath interference, we provide examples of multisnapshot and no-noise scenarios, respectively.
For the multisnapshot scenario, we set $(\phi^{[1]}(1),\phi^{[2]}(1)) = (0,0)$ and $(\phi^{[1]}(t),\phi^{[2]}(t)) = (\pi,0)$, $t=2,3,4,5$.
Fig.\,\ref{fig:Fig-orgAoA}(b) illustrates the corresponding MAE results.
The low MAE area is larger than that in Fig.\,\ref{fig:Fig-orgAoA}(a), indicating the substantial improvement in AoA accuracy due to the noise reduction by multiple snapshots.
If the observation time slots can be sufficiently long, the noise effect can be completely removed while the multipath effect still exists.
Fig.\,\ref{fig:Fig-orgAoA}(c) illustrates the corresponding MAE results under no noise scenario with a single snapshot.
The red area then illustrates the area that suffers from multipath interference.

Fig.\,\ref{fig:non-orgAoA} show the MAE of AoA estimation for the multilabeling mode under different numbers of snapshots.
For $T=3$, we set $(\phi^{[1]}(1),\phi^{[2]}(1)) = (0,0)$, $(\phi^{[1]}(2),\phi^{[2]}(2)) = (2\pi/3,4\pi/3)$, and $(\phi_1(3),\phi_2(3)) = (4\pi/3,8\pi/3)$.
Through path extraction and association, we can directly estimate the AoA from RIS1 and RIS2.
The corresponding MAE results are shown in Fig.\,\ref{fig:non-orgAoA}(a).
Compared with Fig.\,\ref{fig:Fig-orgAoA}(a), the low MAE area (the blue area) is reduced due to the inter-RIS interference.

For multiple snapshots with ${T=9}$, we set $(\phi^{[1]}(1),\phi^{[2]}(1)) = (0,0)$, $(\phi^{[1]}(2j),\phi^{[2]}(2j)) = (2\pi/3,4\pi/3)$, $(\phi^{[1]}(2j+1),\phi^{[2]}(2j+1)) = (4\pi/3,8\pi/3)$, $j=1,\ldots,4$.
In Fig.\,\ref{fig:non-orgAoA}(b), the low MAE area is larger than that in Fig.\,\ref{fig:non-orgAoA}(a) because the noise effect can be largely reduced by the multisnapshot observations.
We collect the MAE results from Points A and B of Fig.\,\ref{fig:Fig-SS} and report a CDF plot for different snapshots in Fig.\,\ref{fig:non-orgAoA-CDF}.
In Point A, the performance of RIS1 is better than that of RIS2 because RIS1 is close to Point A.
Similarly, the performance of RIS2 is better than that of RIS1 in Point B.
At Points A or B, the performance of $T = 9$ is better than $T = 3$.
In Fig.\,\ref{fig:non-orgAoA-CDF}(a), MAE of $90\%$ is less than $10^{\circ}$ for RIS1 of $T = 9$, and MAE of $70\%$ is less than $10^{\circ}$ for RIS1 of $T = 3$.
In Fig.\,\ref{fig:non-orgAoA-CDF}(b), MAE of $95\%$ is less than $10^{\circ}$ for RIS2 of $T = 9$, and MAE of $85\%$ is less than $10^{\circ}$ for RIS2 of $T = 3$.
The channel gain of RIS2 at Point B is larger than RIS1 at Point A. Thus, the performance of RIS2 at Point B is better than that of RIS1 at Point A.

The above results show that the noise effect on multipath labeling can be mitigated by long snapshots. However, long snapshots cannot solve the multipath and the inter-RIS interference effects.
Therefore, we suggest using the single-labeling mode to remove the inter-RIS interference.

\subsection*{C. Localization}
\begin{figure}[!ht]
\begin{flushright}
\setlength{\abovecaptionskip}{-1.0pt}
\begin{center}
\resizebox{2.75 in}{!}{%
\includegraphics*{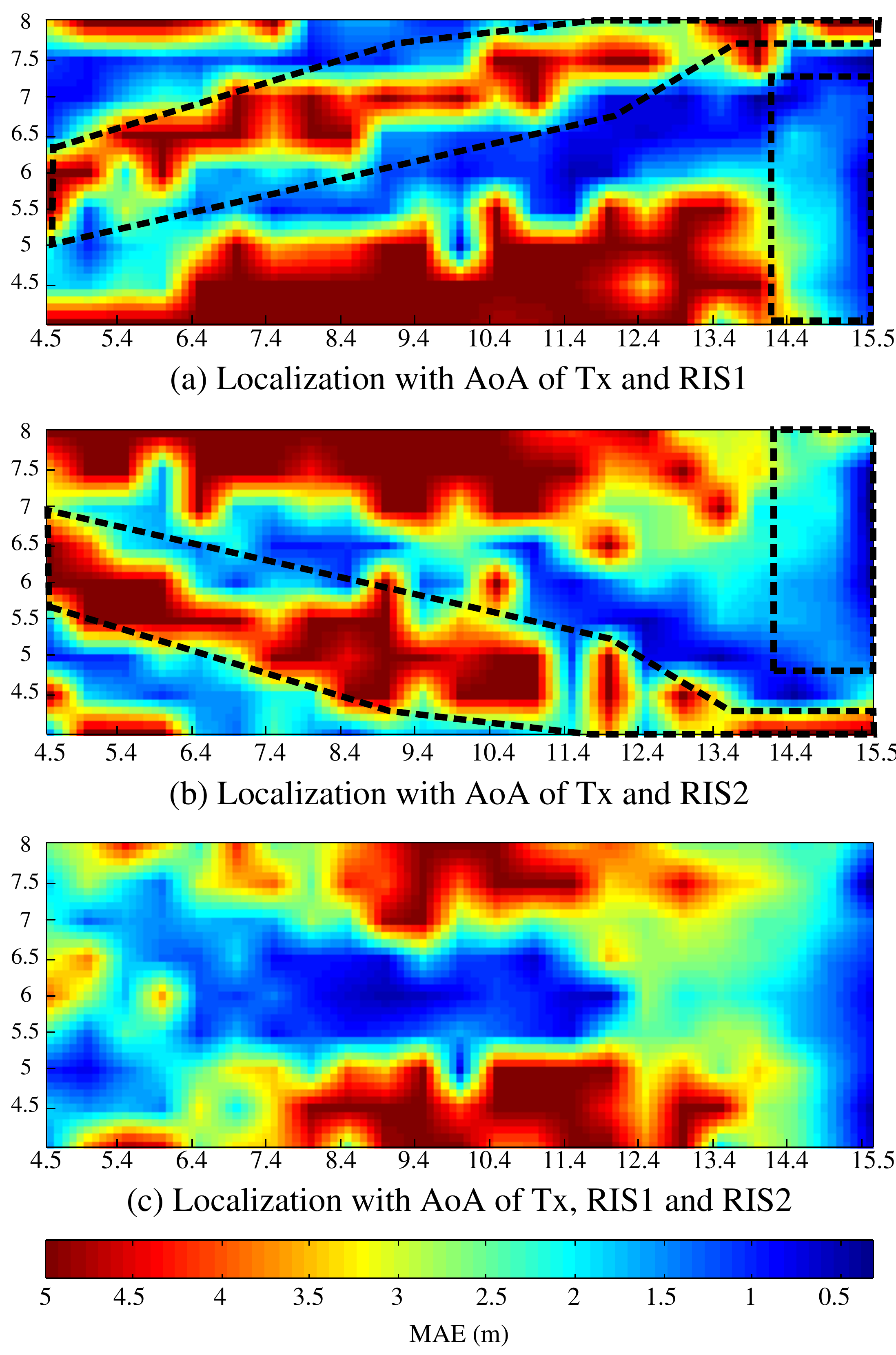} }%
\caption{MAE of localization under the single-labeling mode with $T=9$ for different locations of Rx on the basis of AoAs from (a) Tx and RIS1, (b) Tx and RIS2, (c) Tx, RIS1, and RIS2. Black boxes represent the exception areas.}\label{fig:Fig_localization}
\end{center}\vspace{5.0pt}
\resizebox{3.05 in}{!}{%
\includegraphics*{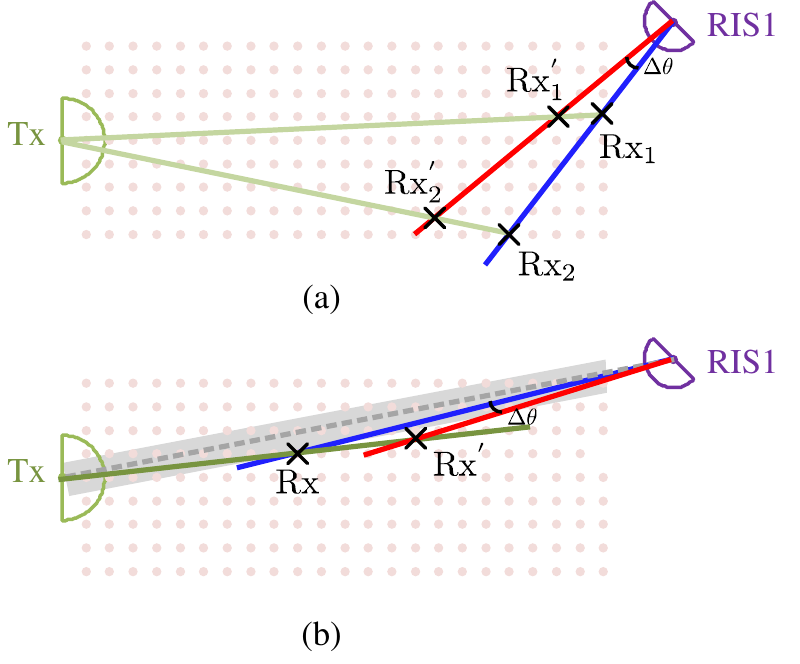} }%
\caption{Localization results analysis for (a) the right side area near RIS1 and (b) the direct line area between Tx and RIS1.
Blue and red lines represent the lines between RIS and Rx on the basis of the correct and estimation AoAs, respectively.
Green lines represent the lines between Tx and Rx. Gray indicates the area close to the direct line between Tx and RIS1.
}\label{fig:Fig_loc_e}
\end{flushright}
\end{figure}

As an application, we consider the localization of Rx by using the estimated AoAs. We assume that the positions of Tx and RISs are known. 
Then, the Rx localization can be estimated by the closest point among the AoA intersections for the direct paths of RIS-to-Rx and Tx-to-Rx. 
The environment is the same as Fig. \ref{fig:Fig-SS}. We use the single-labeling mode with multiple snapshots with $T=9$. In this mode, the AoA of RISs can be obtained without inter-RIS interferences.
We use the MAE of Rx position to indicate the performance of the localization results.
Figs.\,\ref{fig:Fig_localization}(a)-(c) show the corresponding results with RIS1, RIS2, and both assistances, respectively.  
Note that one Tx without RIS assistance cannot perform localization through AoA.  
Therefore, we only show the localization results with RIS assistance.

The low MAE area (the blue area) in Fig.\,\ref{fig:Fig_localization}(a) is similar to that in Fig.\,\ref{fig:Fig-orgAoA}(b) except for 1) the right side area near RIS1 and 2) the direct line area between Tx and RIS1.
The two exceptions reverse the results of Fig.\,\ref{fig:Fig-orgAoA}(b). For example, the right side area near RIS1 of Fig.\,\ref{fig:Fig-orgAoA}(b) is poor in AoA estimation while surprisingly good in localization in Fig.\,\ref{fig:Fig_localization}(a).
To better understand the reasons for the two exceptions, we illustrate through the examples in Figs. \ref{fig:Fig_loc_e}(a) and \ref{fig:Fig_loc_e}(b), respectively. First, let us assume that the estimation quality of the AoA for the direct path of Tx-to-Rx is precise because the power of Tx is relatively large. Then, we consider that the AoA estimations from RIS1 have error $\Delta\theta$. With the error, the localizations of ${\rm Rx}_1$ and ${\rm Rx}_2$ are changed to ${\rm Rx}_1'$ and ${\rm Rx}_2'$, respectively.
The two locations have quite different localization errors even they have the same AoA error. Clearly, the Rx located close to RIS1 has a smaller localization error than that far from RIS1.   
Next, we consider the reason for another exception. Fig.\,\ref{fig:Fig_loc_e}(b) illustrates that a small AoA error results in a significant localization error when Rx is on the direct line area between Tx and RIS1.
The localization error is sensitive to the AoA error in the gray area, which results in poor localization results in Fig.\,\ref{fig:Fig_localization}(a).
The same characteristics can be observed from Fig.\,\ref{fig:Fig_localization}(b), where RIS2 is used as the assistance. 
When RIS1 and RIS2 are used, Fig. \ref{fig:Fig_localization}(c) shows that multiple RISs extend the sensing area. 

\section*{ V. Conclusion}
This study presents multipath labeling, which consists of channel flipping, parameter extraction, and
RIS association. The simulation results reveal that the labeled path's parameters can be accurately extracted through multiple snapshots even under the sub 6GHz channels with a severe multipath effect.
With the assistance of RIS, the localization can be achieved through a single Tx, and its sensing area can be extended by multiple RISs. The feasibility of multipath labeling can open the door for more efforts in environmental sensing direction.

{\renewcommand{\baselinestretch}{0.98}

}

\end{document}